\documentclass[conference]{IEEEtran}
\IEEEoverridecommandlockouts
\usepackage{cite}
\usepackage{amsmath,amssymb,amsfonts}
\usepackage{algorithmic}
\usepackage{textcomp}
\usepackage{xcolor}
\usepackage[pdftex]{graphicx}
\usepackage{epstopdf}
\usepackage{bm}
\usepackage{caption}
\usepackage{subfigure}
\usepackage{subfigmat}
\usepackage{comment}

\usepackage{amsmath,amssymb,enumerate} 
\usepackage{bm} 
\usepackage{indentfirst} 
\usepackage{empheq} 
\usepackage{here} 

\def\BibTeX{{\rm B\kern-.05em{\sc i\kern-.025em b}\kern-.08em
    T\kern-.1667em\lower.7ex\hbox{E}\kern-.125emX}}
\begin{document}

\title{A Model of Polarization on Social Media Caused by Empathy and Repulsion}

\author{\IEEEauthorblockN{Naoki  Hirakura}
    \IEEEauthorblockA{
        \textit{Tokyo Metropolitan University} \\
Tokyo 191--0065, Japan \\
hirakura-naoki@ed.tmu.ac.jp}
\and
\IEEEauthorblockN{Masaki Aida}
    \IEEEauthorblockA{
        \textit{Tokyo Metropolitan University} \\
Tokyo 191--0065, Japan \\
aida@tmu.ac.jp}
\and
\IEEEauthorblockN{Konosuke Kawashima}
    \IEEEauthorblockA{
        \textit{Tokyo Metropolitan University} \\
Tokyo 191--0065, Japan \\
k-kawa@tmu.ac.jp}
}

\maketitle

\begin{abstract}
In recent years, the ease with which social media can be accessed has led to the unexpected problem of a shrinkage in information sources.
This phenomenon is caused by a system that facilitates the connection of people with similar ideas and recommendation systems.
Bias in the selection of information sources promotes polarization that divides people into multiple groups with opposing views and creates conflicts between opposing groups.
This paper elucidates the mechanism of polarization by proposing a model of opinion formation in social media that considers users' reactions of empathy and repulsion.
Based on the idea that opinion neutrality is only relative, this model offers a novel technology for dealing with polarization.
\end{abstract}


\section{Introduction}
Today, social media services such as Twitter and Instagram have become widely adopted and are playing a significant role in information distribution.
Unlike traditional mass media, social media allows users to efficiently focus on views that trigger their interest.
Paradoxically, however, its very power raises the problem of excluding most information sources.
This is because of the feature that allows users to choose the accounts they chose to follow and the existence of push-style recommendation systems.
It is feared that social media will lead to polarization because it makes it easier to connect people with similar opinions which biases the information distribution~\cite{sunstein2018republic}.
Polarization is the division of opinion on a topic into basically two diametrically opposing positions.

Polarization tends to emerge when users express their opinions within a shuttered community; the opinions expressed are almost always seen as positive, with negative opinions being drowned or pushed out.
As a result, a special understanding that is common only within the community becomes strengthened, and a gap forms between the thoughts and values of people outside the community.
This reinforcement of belief within a community is called the echo chamber phenomenon.
Due to such bias in information distribution, slander, and defamation between communities that have different opinions become more frequent.
To prevent these social problems, we should elucidate the mechanism of polarization on social media.

The basic reactions of users to posts on social media are empathy and repulsion.
The repulsive reaction is called the backfire effect in psychology~\cite{nyhan2010corrections}, and one report has confirmed the backfire effect of social media use~\cite{bail2018exposure}.

Reference~\cite{chen2019opinion} proposed a model of opinion formation that incorporates the backfire effect.
In this model, polarization does not occur automatically in the distribution of biased opinions because the absolute origin is determined by the quantitative representation of opinions.
This model corresponds to a situation where objective "neutral" positions are predetermined with regard to the distribution of opinions. Polarization is assumed to occur between users whose opinions are on the opposite sides of the neutral position.

In this paper, we propose a model that users are influenced by other users via social media such that their opinions change.
A feature of this model is the origin of the quantitative representations of opinions is not necessarily objective neutrality.
This feature reflects our view that opinion neutrality is relative.
That is, polarization is generated from relative differences in opinion, but not from the predetermined standard of neutrality.
Also, the proposed model is designed to yield different characteristics based on differences in opinion.
There are two types of specific influence: one is a strong empathy for posts that are close to one's own opinion and dismissal of different opinions.
The other is a strong repulsion for posts that differ from their own opinion and a ignoring of opinions that are close to their own.

The structure of this paper is as follows.
In Section 2, we propose a model of opinion formation in which users show empathetic and repulsive reactions to posts on social media.
In Section 3, we evaluate the model proposed in Section 2 and clarify the behaviors of the model for given parameters.
Section 4 compares the proposed method with previous works and shows the advantages of the proposed method.
We conclude in Section 5.

\section{Modeling Polarization by Empathy and Repulsion}
\label{model}
We model situations in which a user's posts on social media influence the opinions of other users; two kinds of reactions are considered, empathy and repulsion.
When empathy occurs, users strongly empathize with posts that are close to their opinions and dismiss posts that express opinions very different from their own.
In the case of repulsion, users are repulsed more by posts that are more distant from their opinions, and less likely to be repulsed by posts that are closer to their own opinions.
The following are the rules for changing opinions based on these reactions.

Consider a social network with $N$ users.
User $i$ $(i=1,\,\dots,\,N)$ is deemed to have opinion value $o_i(t) \in [-1,1]$; the opinion value $o_i^p(t)$ of user $i$ for a post at time $t$ is $o_i^p(t)=o_i(t)$ reflects the user's opinion at that time.
We assume that the opinion of user $i$ is affected by concurrent posts.
If user $j$ is the author of the latest post seen by user $i$, then the opinion value of user $i$ at time $t$ changes due to $o_j^p(t^-)$.
The opinion value of user $i$ is given by $o_i(t) = o_i(t^-) + f(o_i(t^-), \, o_j^p(t^-))$.
The second term on the right side will be explained later.

When user $i$ sees the latest post of user $j$, user $i$ develops either an empathetic or repulsive response with the probability of $p_i$ or $1-p_i$, respectively.
We assume that the magnitude of empathy and repulsion depends on the difference in opinion values.

First, consider the case of empathy.
User $i$ should show a stronger empathetic reaction to posts whose opinion values are close to their own, and weaker empathy when the opinion values are very different.
Based on this, we introduce function $f(o_i(t^-), \, o_j^p(t^-))$, which represents the change in the opinion value, as 
\begin{align}
\Delta \,c\,k\, \exp(-k\,|\Delta|). 
\label{empathy}
\end{align}
Note that $\Delta=o_j^p(t^-)-o_i(t^-)$ is the difference between the opinion value of the post seen by user $i$ in latest and the user's opinion value.
Parameter $k (> 0)$ is a constant that determines the decay in the intensity of empathy with respect to the absolute value of the difference of opinion values $|\Delta|$, and $c$ is a parameter that is adjusted to be $o_i(t)\in[-1,1]$, which satisfies $ck\leq 1$.

Next, consider the case of repulsion.
In this case, it is necessary to develop a stronger repulsive reaction to the opinion values that are very different from their own opinion values, and the repulsion should be weaker when the opinion values are close.
To express the repulsive reaction in the same form as in the case of empathy, we impose a periodic boundary condition on the two ends of $+1$ and $-1$ to make the difference between the opinion value of a post and the user's opinion $|\Delta'|=2-|\Delta|$; the smaller the difference $|\Delta'|$, the larger the repulsive reaction.
Based on this, we design function $f(o_i(t^-), \, o_j^p(t^-))$, which represents the change in opinion value, as 
\begin{align}
&\left(1-o_{i}(t^-)\right) c \, k \, \exp\! \left(-k\,|\Delta^{\prime}|\right),\quad (\Delta\leq0), 
\label{repulsion_negative}
\\
&-\left(1+o_{i}(t^-)\right) c \, k \, \exp\! \left(-k\,|\Delta^{\prime}|\right),\quad (\Delta>0). 
\label{repulsion_positive}
\end{align}

\begin{figure*}
  \centering
  \subfigure[$\Delta \leq 0$]{
    \includegraphics[width=0.65\linewidth]{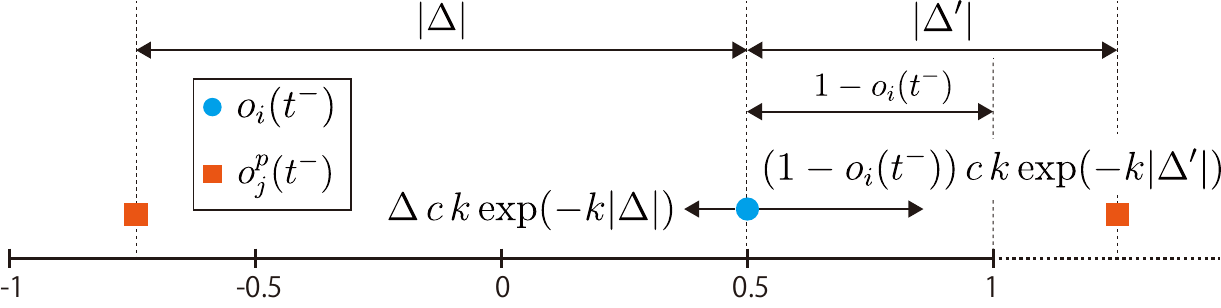}
    \label{opinion_change_a}
  }
  \subfigure[$\Delta > 0$]{
    \includegraphics[width=0.65\linewidth]{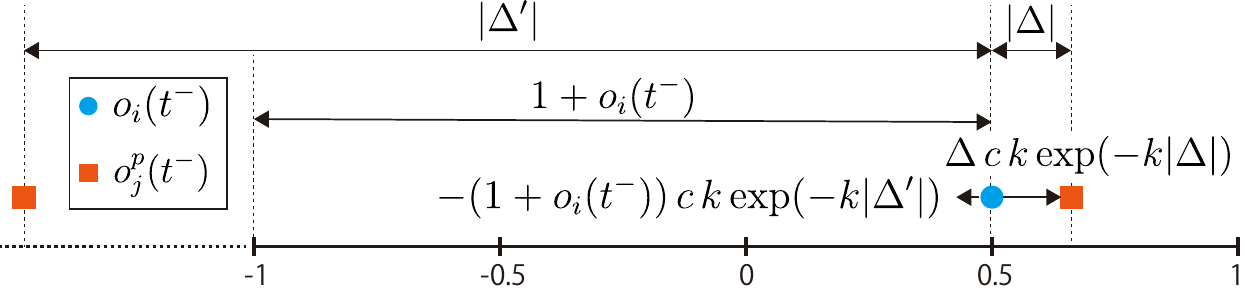}
    \label{opinion_change_b}
  }
  \caption{Intuitive images of opinion value change.}
  \label{opinion_change}
\end{figure*}

Figure \ref{opinion_change} illuminates the concrete actions of expressions \eqref{empathy}, \eqref{repulsion_negative}, and \eqref{repulsion_positive}.

The expression \eqref{empathy} shows the change in the opinion value in the case of an empathy reaction; it indicates that the user's opinion value approaches the opinion value of the post by an amount proportional to the difference, $c\,k\, \mathrm{exp}(-k\,|\Delta|)$.
In this case, the positive or negative value of $\Delta$ corresponds to the direction of the movement in the user's opinion to be closer to the opinion of the post as shown in the negative direction of Fig.~\ref{opinion_change_a} and the positive direction of Fig.~\ref{opinion_change_b}.

Figure \ref{opinion_change_a} is the case of $\Delta\leq 0$.
The arrow pointing in the negative direction from the blue circle representing the user's opinion value indicates that the user's opinion value approaches the opinion value of a post by the proportion of $c\,k\, \exp(-k\,|\Delta|)$ of the difference $|\Delta|$.
On the other hand, Fig.~\ref{opinion_change_b} is the case of $\Delta > 0$, where the sign of $\Delta$ denotes a positive change in opinion value in the opposite direction to that of figure \ref{opinion_change_a}.

The expressions \eqref{repulsion_negative} and \eqref{repulsion_positive} show the change in opinion value under a repulsion reaction.
First, we will discuss the case of $\Delta \leq 0$.
As shown in Fig.~\ref{opinion_change_a}, the opinion value of user $i$ increases by the proportion $c\, k\, \exp(-k\,|\Delta^{\prime}|)$ of the difference $|1-o_i(t^-)|$ between the opinion value of user $i$ and the most extreme opinion value.
In this case, the amount of change in the opinion value becomes the expression \eqref{repulsion_negative} and the user's opinion value moves away from the posted opinion value.
We now discuss the case of $\Delta > 0$.
As shown in Fig.~\ref{opinion_change_b}, the opinion of user $i$ decreases by the proportion of $-(1+o_i(t^-))$ of the difference $-(1+o_i(t^-))$ between the opinion value $o_i(t^-)$ of user $i$ and the most extreme opinion $-1$.
In this case, the amount of change in the opinion value is given by the expression \eqref{repulsion_positive}, where the user's opinion value moves away from the posted opinion value.

The parameter $k$ $(> 0)$ represents the decay rate of the strength of the effect, and the value of $k$ characterizes the rate of change in the strength of the effect based on the difference in opinion values.
Parameter $k$ allows us to design opinion value change rules where users show a stronger empathetic reaction to posts with close opinion values and a stronger repulsive reaction to posts with distant opinion values.

\section{Experimental evaluations of the proposed model}
\subsection{Preparation for evaluation experiments}
We explain how to generate a sequence of posts and an index of polarization for the experiments conducted on the proposed model.

\subsubsection{Generating a sequence of posts}\label{post_generating}
Here, we generate artificial data using the multivariate Hawkes process~\cite{hawkes1971spectra}.
The multivariate Hawkes process is a point process in which multiple processes mutually excite the occurrence of events.
To use this for the generation of user post sequences on social media, we assume that each process is considered to be a user and each event is a post to social media and that posts are promoted between connected users.

The posting rate $\lambda_i$ of user $i$ $(i=1,\,\dots,\,N)$ is expressed as follows.
\begin{align}
  \lambda_{i}(t)=\mu_{i}+\sum_{j \in \partial i} \sum_{t_{h} \in H_{j}} \alpha_{j i} \, \mathrm{e}^{-\beta_{j i}\left(t-t_{h}\right)},
  \label{hawkes}
\end{align}
where $\mu_i$ is the base rate, $\partial i$ is the set of neighbors of user $i$, $H_j$ is the set of past posting times of user $j$, and $t_h$ is its element.
$\alpha_{ji}$ and $\beta_{ji}$ $(j=1,\,\dots,\,N, \, i=1,\,\dots,\,N)$ represent the rate jump and decay rate for user $i$ due to user $j$'s posts, respectively.
Thus, the posting rate of user $i$ is determined by the influence of the user-specific base rate and the previous posts of neighboring users.

We generate a sequence of posts using the multivariate Hawkes process.
First, a social network with $10$ nodes is created by the Barab{\'a}si-Albert model (hereinafter referred to as the BA model)~\cite{Barabasi509}.
In eq.~\eqref{hawkes}, $\alpha_{ji}$ and $\beta_{ji}$ are given as uniform random numbers in the range of $[0,1]$ for each combination of $j$ and $i$ $(j,\,i =1,\dots,N)$.
The base rate $\mu_i$ is given as a uniform random number in the range of $[0,1]$.
We evaluate the proposed model by generating a sequence of posts in the observation period $[t_1, t_1+2]$ with the first post occurrence time $t_1$.

\subsubsection{Index of polarization}
We introduce the index~\cite{esteban1994measurement} for frequency distribution $(\bm{\pi},\bm{y}) = (\pi_1,\,\dots,\pi_n \,;\, y_1,\,\dots,\,y_n)$ as the index of polarization: 
\begin{align}
  P=K \sum_{i=i}^{n} \sum_{j=1}^{n} \pi_{i}^{1+\theta} \, \pi_{j} \, \left|y_{i}-y_{j}\right|, 
  \label{er_index}
\end{align}
where $n$ is the number of classes that equally divide the opinion value space $[-1,1]$, $y_i$ is the $i$-th class value from the bottom, $\pi_i$ is the number of users belonging to the $i$-th class, $K$ is a parameter for normalization, and $\theta$ is a parameter called the polarization sensitivity, which takes a value in the range of $(0,\theta^{*} \simeq 1.6)$.
According to \cite{esteban1994measurement}, the stronger the sense a user has to belonging to a class and the greater the degree of hostility toward another class, the greater is the degree of polarization.
The index of polarization \eqref{er_index} is designed to satisfy the appropriate axioms for a valid index of polarization.
For example, the maximum value is taken if half of the users belong to the lowest and half to the highest class, respectively, and the minimum value is taken if all users belong to the same class.

\subsection{Parameter characteristics of the proposed model}
We conduct evaluation experiments to elucidate the characteristics of the parameters of the proposed model.
The experimental conditions are as follows.
The initial opinion is given as a uniform random number of $[-1,1]$.
We apply the opinion value change rule to the sequence of contributions generated in Sect.~\ref{post_generating}, with the probability of an empathetic response and the probability of a repulsive reaction being $p_i=p$ and $1-p_i=1-p$, respectively, for all $i$.

\begin{figure}[tb]
  \begin{center}
    \subfigure[$p=0.9$, $k=2$]{
      \centering
      \includegraphics[width=0.7\linewidth]{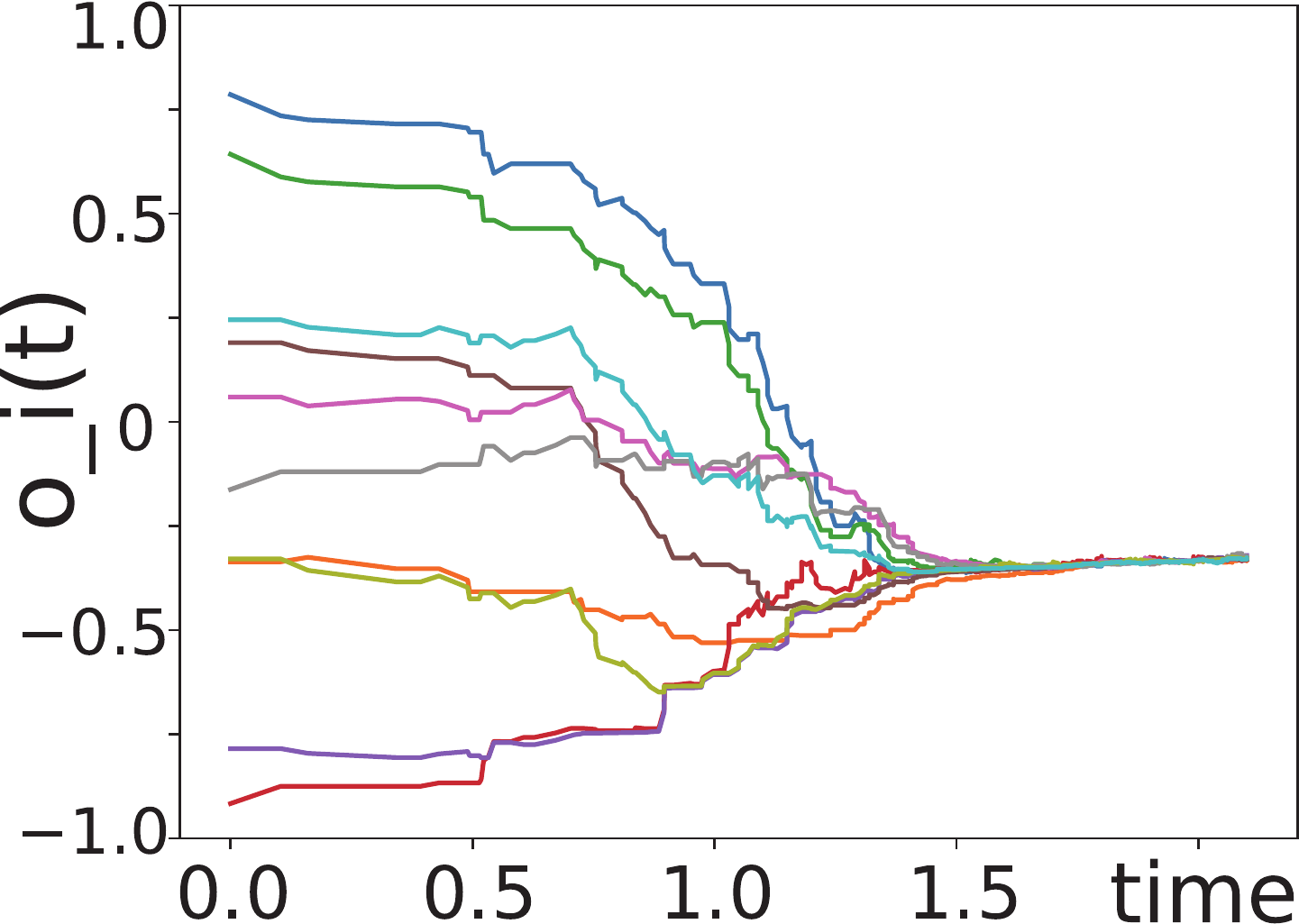}
      \label{fig:p0.9_k2}
    }
    \subfigure[$p=0.1$, $k=2$]{
      \centering
      \includegraphics[width=0.7\linewidth]{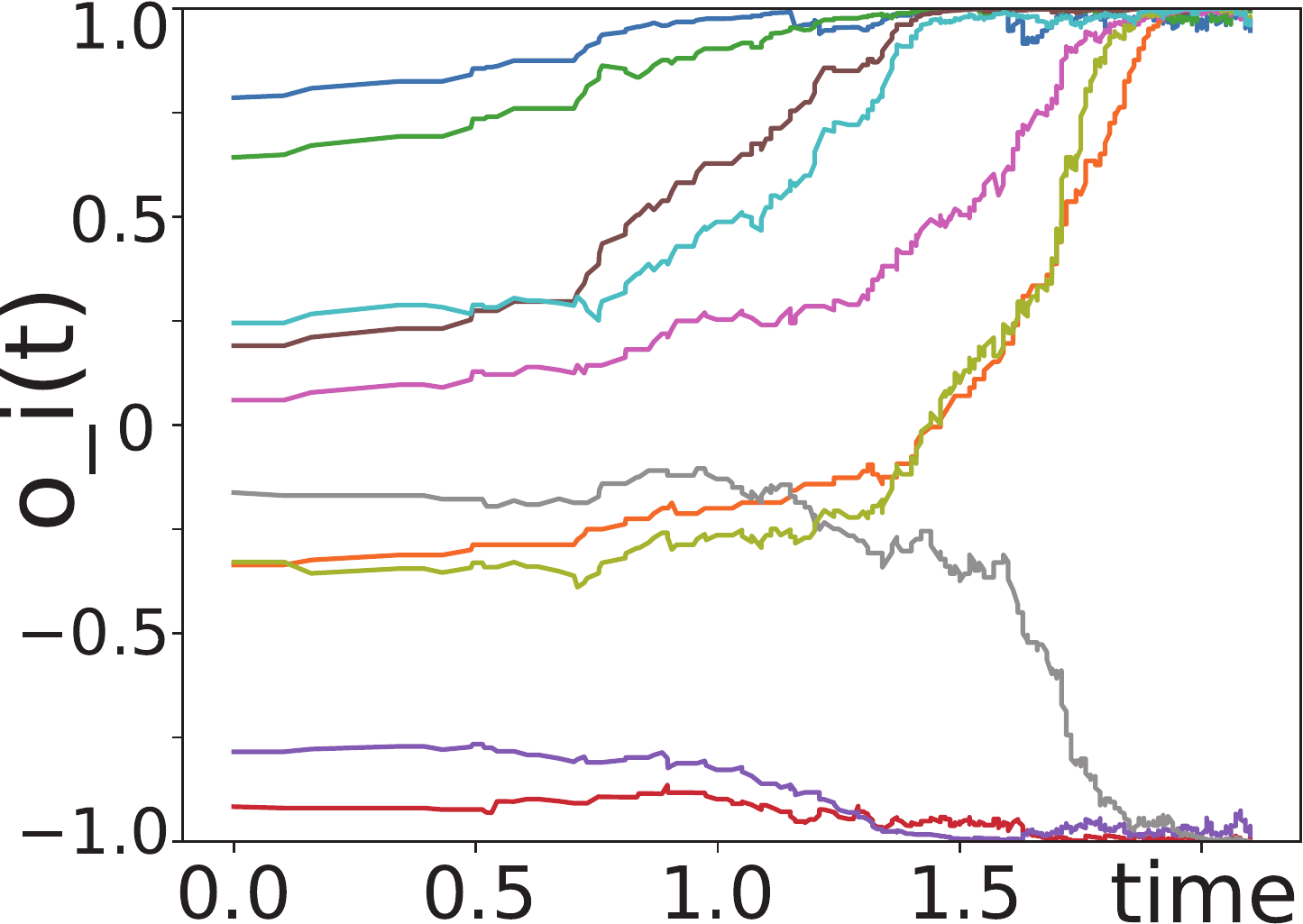}
      \label{fig:p0.1_k2}
    }
    \subfigure[$p=0.9$, $k=10$]{
      \centering
      \includegraphics[width=0.7\linewidth]{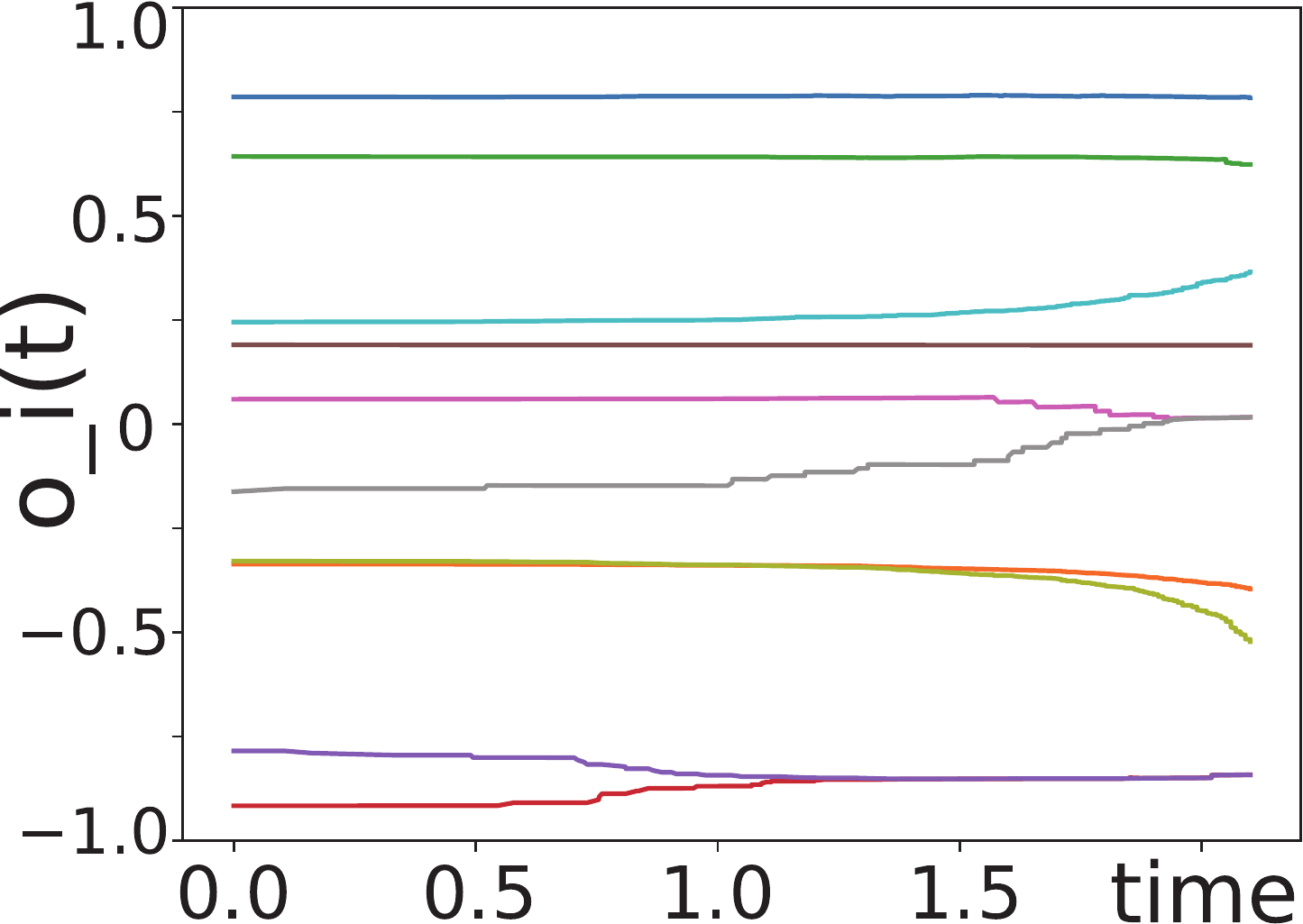}
      \label{fig:p0.9_k10}
    }
    \caption{Examples of opinion value change.}
    \label{fig:ex_opinion_change}
  \end{center}
\end{figure}

Three typical simulation results with different parameters for the opinion change using the proposed model are shown in Figs.~\ref{fig:p0.9_k2}, \ref{fig:p0.1_k2} and \ref{fig:p0.9_k10}.
Parameter $k=2$ is fixed for the cases of Figs.~\ref{fig:p0.9_k2} and \ref{fig:p0.1_k2}, and the probability of empathy $p$ is set to $p=0.9$ and $p=0.1$, respectively.
The difference in the probability of empathy reaction $p$ causes a consensus of opinions in Fig.~\ref{fig:p0.9_k2} and a polarization of opinions in Fig.~\ref{fig:p0.1_k2}.
Figure~\ref{fig:p0.9_k10} shows the case of $k=10$.
Since the influence of other users' posts is smaller than when $k=2$, the opinion values of each user do not change much from their initial values.

\begin{figure}[tb]
  \centering
  \includegraphics[width=\linewidth]{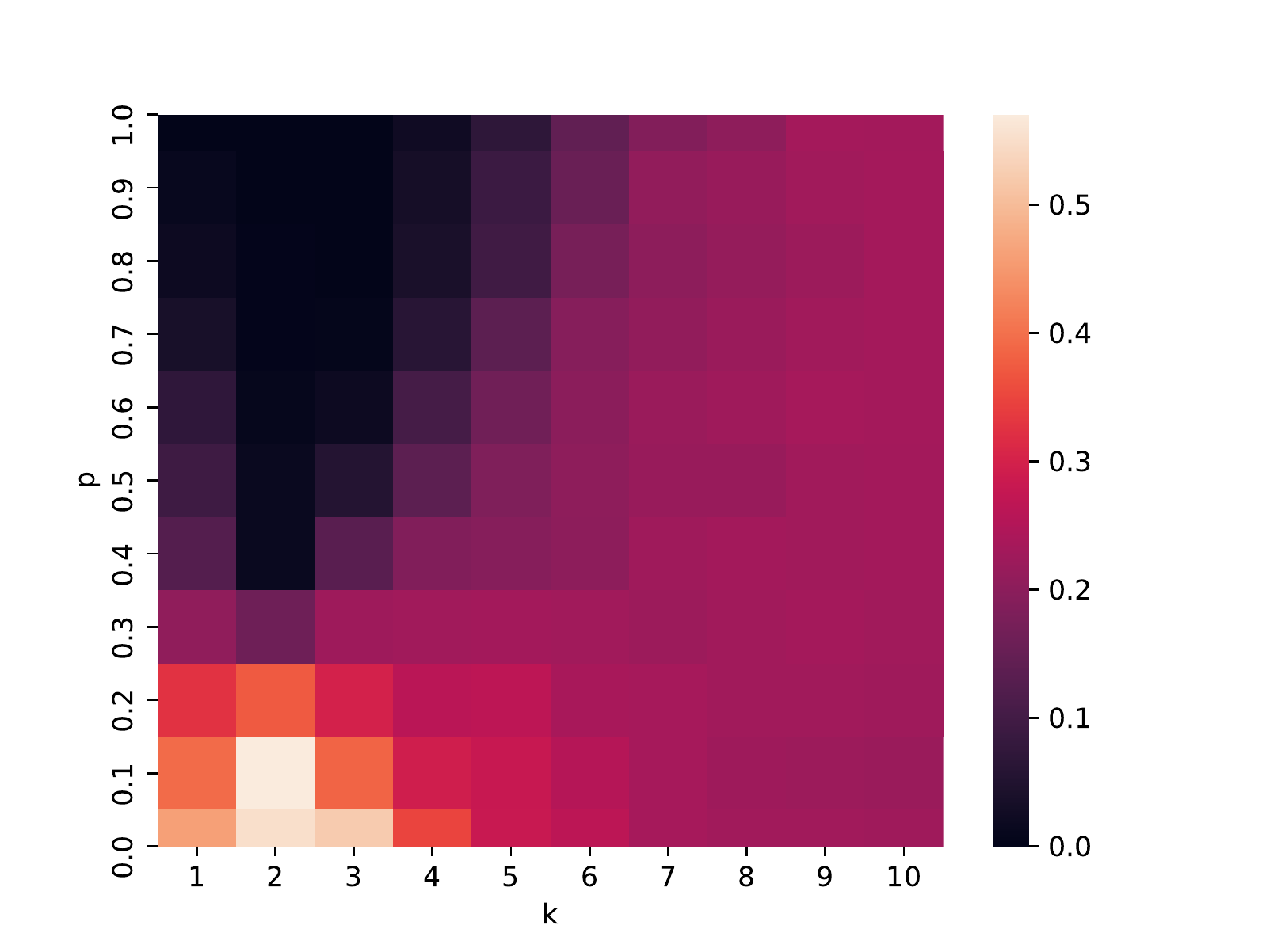}
  \caption{Parameter characteristics of the proposed model.}
  \label{fig:pol}
\end{figure}

We then evaluate the polarization of the final state at the final simulation time using the index of polarization \eqref{er_index}, for each combination of parameters $p$ and $k$.
The normalization parameters are set to $K=\left(\sum_{i=1}^{n}\pi_i\right)^{-(2+\theta)}$, the polarization sensitivity is set to $K=0.5$, and the number of classes is set to $n=10$.
Thus the possible values of the polarization index are approximately $[0,0.62]$.
Assuming $c=0.05$, Fig.~\ref{fig:pol} shows the average of $10$ simulations for each combination of $p=0,\,0.1,\,\dots,\,1$ and $k=1,\,\dots,\,10$ with different initial opinion values.
According to Fig. \ref{fig:pol}, the polarization at the end of the simulation can be divided into three main types: concentrated into one opinion, polarized into two opinions, and little change in the opinions of each user.
Opinions can be concentrated into one opinion when roughly $p \geq 0.4$ and $1 \leq k \leq 5$.
This tendency occurred because they were more likely to show an empathic reaction even when the difference in opinion values was only slight.
The polarization into two opinions occurs when $p < 0.4$ and $1\leq k\leq 4$.
This tendency occurred because they were more likely to show a repulsive reaction even when the difference in opinion values was only slight.
The opinion of each user hardly changes when $k \geq 5$, regardless of the value of $p$.
As $k$ increases, it is due to the fact that we empathize only with those opinions that are very close and reject only those that are far enough away.

\section{Comparison with previous research}
We compare our model with the opinion formation model, which takes into account the reactions corresponding to empathy and repulsion.
As an example, we take the BEBA model \cite{chen2019opinion}.
The BEBA model is an expansion of classical opinion formation model, DeGroot model~\cite{degroot1974reaching}; it contains biased assimilation and the backfire effect.
Biased assimilation means that we highly evaluate information that is consistent with our original opinion, and the backfire effect means that when we are exposed to an idea that is different from our own, we come to strongly believe our original opinion.

The BEBA model is briefly described as follows.  
First, consider a social network with $N$ users.
User $i$ $(i=1,\,\dots,\,N)$ has opinion value $y_i(t)\in[-1,1]$ at time $t$.
For adjacent users $i$ and $j$, if the opinion values $y_i(t)$ and $y_j(t)$ are the same sign, biased assimilation controls, and if they are different signs, backfire effect dominates.
This means that the opinion value $0$ is neutral and users have different opinions depending on whether they have positive or negative opinion values.
On the other hand, in the proposed model, the amount of change in the opinion value is determined based on the difference between the post's and user's opinion values.
To clarify the difference between the two models, we give all the initial opinion values with the same sign in a comparative evaluation.

\begin{figure}[tb]
  \begin{center}
  \subfigure[BEBA model]{
    \includegraphics[width=0.7\linewidth]{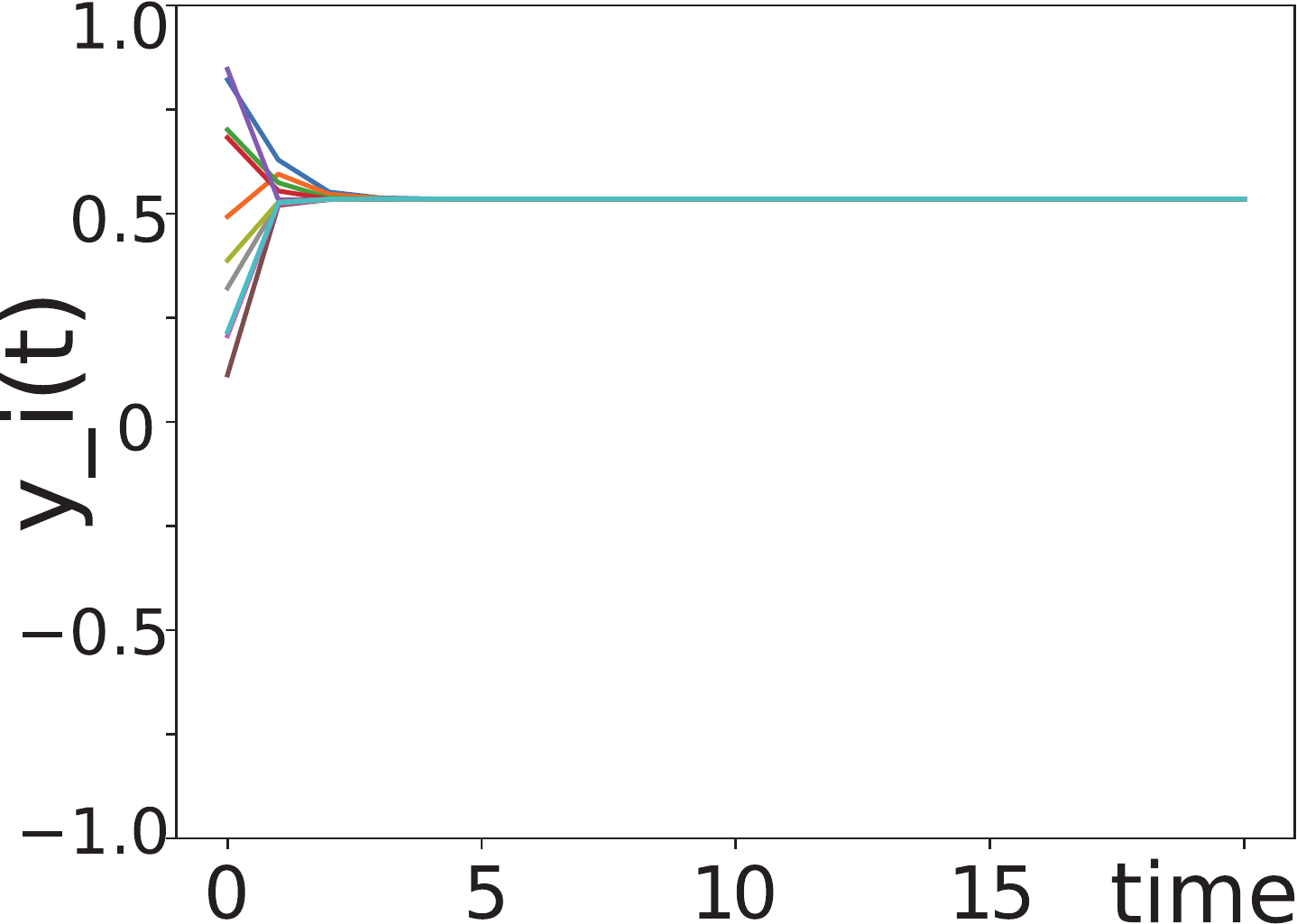}
    \label{fig:beba_positive}
  }
  \subfigure[Proposed model]{
    \includegraphics[width=0.7\linewidth]{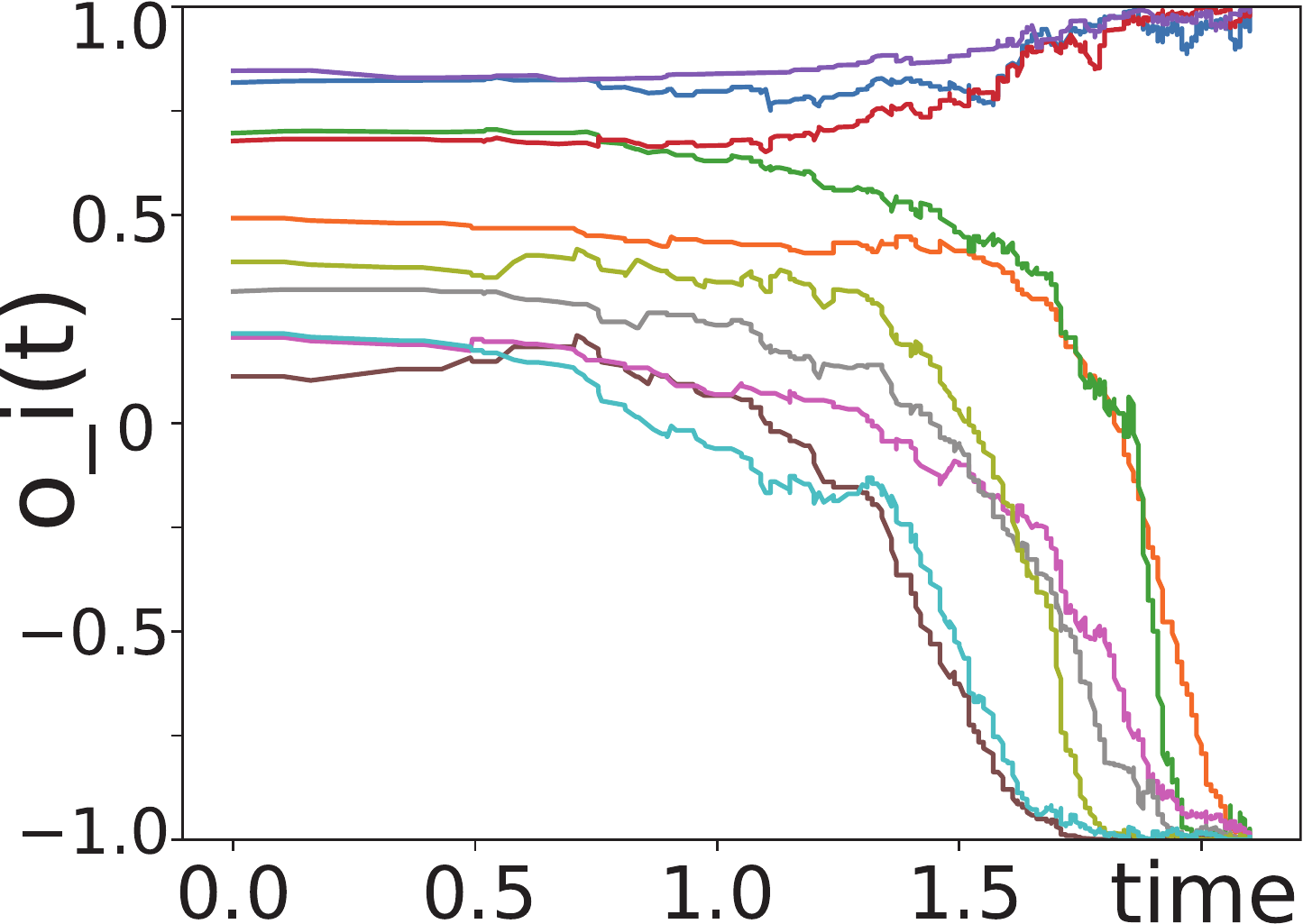}
    \label{fig:proposal_positive}
  }
  \end{center}
  \caption{Change in opinion values when initial opinion values given to all users have the same sign.}
  \label{fig:compare_model}
\end{figure}

First, we describe the results of the experiments on the BEBA model.
Figure \ref{fig:beba_positive} shows the change in opinion values yielded by the BEBA model.
Since the initial opinion values with the same sign were given to all nodes, the backfire effect was quiescent, and the opinions became concentrated at the end of the simulation.
Next, we describe the results of the experiments on the proposed model.
Figure \ref{fig:proposal_positive} shows the change in opinion values for $p=0.1$ and $k=2$ yielded by the proposed model.
Although the initial opinion values with the same sign were given to all nodes, the amount of change in the opinion values was determined based on the difference in the opinion values, so that a repulsion occurred and polarization commenced.
Depending on the choice of parameters, the proposed model was found to be able to handle cases of polarization in which repulsive reactions were active even with a biased opinion value distribution.

\section{Conclusion}
In this paper, we proposed a model of social media users' opinion formation to elucidate the mechanism of opinion polarization.
The model is characterized by two types of reactions: empathy and repulsion, with different strengths of influence being created by differences in user's opinion values.
It also reflects the idea that relative differences in opinion determine the neutrality of opinion.
In particular, since our idea of opinion neutrality dispenses with the concept of objective opinion neutrality, it has the advantage of being able to respond flexibly to a biased distribution of opinion values.
These features allow a wide range of user characteristics to be appropriately modeled so as to well cover the phenomenon of polarization.
The dependence of the change in the opinion value on the initial values of the opinions will be examined in the future.

\newpage 

\section*{Acknowledgement}
This research was supported by Grant-in-Aid for Scientific Research 19H04096 and 20H04179 from the Japan Society for the Promotion of Science (JSPS).


\end{document}